\documentclass[fleqn]{article}
\usepackage{cite}
\usepackage{amsmath}
\usepackage{graphicx}
\usepackage[mathscr]{eucal}

\graphicspath{%
    {converted_graphics/}
    {./}
}
\begin{document}
\def\vect#1{{\bf #1}}
\def\chir{\raisebox{2pt}{$\chi$}}  
\def\sigmagtm{<\kern-3pt\boldsymbol\sigma\kern-4pt>}
\def\nilsona{\mbox{\large{$\displaystyle a$}}}
\def\coefc{\mbox{\large{$\displaystyle c$}}}

\title{\Large\bf Super-allowed beta-decay rates in $1d_{5/2}$ shell in Coriolis coupling model}

\author{\large M. Sultan Parvez$^1$ and F. Bary Malik$^{2,3}$}
\date{}

\maketitle
{\small
\noindent $^1$Department of Mathematics and Physical Sciences,
 Louisiana State University, 

\qquad Alexandria, Louisiana 71302, U.S.A\\
 $^2$Physics Department, Southern Illinois University, Carbondale, 

\qquad Illinois 62901, U.S.A\\
 $^3$Physics Department, Washington University, St. Louis, 
Missouri, 63130, U.S.A}

\vskip3\baselineskip
\noindent The expression for super-allowed beta-decay transition rates have been derived within the context of Coriolis coupling model. The derived expressions, valid for the beta-decay between any two mirror nuclei, has been applied to calculate super-allowed beta-decay transition rates of  $^{21}$Na,  $^{21}$Mg, $^{21}$Al, and  $^{21}$Si. The calculated rates agree well with the data and the calculations done using the shell model with configuration admixture.

\vskip\baselineskip
\noindent PACS Nos: 21.60Ev, 23.40.-S

\section{Introduction}

Both the shell model with configuration admixture\cite{1,2,3} and the Coriolis 
coupling model\cite{4} reasonably account for the low-lying level spectra, 
and electric quadrupole and magnetic di-pole moments of ground states in many 
odd $1d_{3/2}$-nuclei reasonably well. In addition the Landford and Wildenthal's shell 
model calculations\cite{5} with configuration mixing, using the 
Kuo-Brown\cite{3} 
two-body matrix elements with  $^{16}$O as a core, account for the super-allowed 
beta-decay transition rates between a pair of mirror nuclei in many odd nuclei in $1d$-$2s$ shells. However, there has been no beta-decay transition rate calculations within the context of Coriolis coupling model for $1d_{3/2}$  and other nuclei so far. In this paper, we first 
derive the expressions for super-allowed beta-decay transition rates in Coriolis 
coupling model and apply those to calculate super-allowed transition rates between 
the pairs $^{21}_{11}$Na$\rightarrow ^{21}_{10}$Ne,\,  
$^{23}_{12}$Mg$\rightarrow ^{23}_{11}$Na,\,  
$^{25}_{13}$Al$\rightarrow ^{25}_{12}$Mg,\,  and\,  
$^{27}_{14}$Si$\rightarrow ^{27}_{13}$Al. Kim\cite{6} has calculated the 
super-allowed transition rates of  $^{25}$Al$\rightarrow ^{25}$Mg and  
$^{27}$Si$\rightarrow ^{27}$Al using the Nilsson's\cite{7} 
model. His results for $\log\!ft$ are somewhat lower than the observed data. 
The level spectra of these nuclei are reasonably affected by the Coriolis 
coupling. It is, therefore, interesting to investigate the extent to which 
the band mixing influences the results of superallowed beta-decay.

In the next section, we present a brief outline of the Coriolis coupling model 
of\cite{4} and the theoretical derivation of super-allowed transition within the 
context of that model. This is followed by a section dealing with the application 
to these nuclei.

\section{Theory}

{\it A. Outline of the model}\\

\medskip
\noindent The model Hamiltonian of a nucleon moving in an axial symmetric anharmonic potential of a 
rotating nucleus is given by\cite{8,9,10}
\begin{equation}
 H = {\hbar^2 \over 2\mathscr I }\;\, [ \vect{I}^2               
       + \vect{j}^2 - 2( \vect{I}\kern-3pt\cdot\kern-3pt \vect{j} ) ]  
+ H_p + H_c   \label{H}
\end{equation}
where $\vect{I}$  and $\vect{j}$ are, respectively, the total spin of the body and 
angular momentum of a nucleon, usually the valence one, $\mathscr I$ is the moment of inertia 
of the body, $H_p$ and $H_c$  are, respectively, the intrinsic Hamiltonian governing the 
motion of the nucleon in consideration and the other nucleons in the core in 
body-fixed coordinate  system denoted by primes. The explicit form of $H_c$  is usually not 
considered but $H_p$ is given by the following Nilsson Hamiltonian\cite{7}
\begin{equation}
\label{Hp}
H_p = - {\hbar^2 \over 2\mu} \nabla^\prime{}^2        
            + {\mu \over 2} [ \omega_\rho^2
           (x^\prime{}^2 + y^\prime{}^2)
           + \omega^2_z z^\prime{}^2
          ] + C\,  \vect{l\kern-3pt\cdot\kern-3pt  s} + 
              D\,  \vect{l\!\cdot\kern-1.5pt l}
\end{equation}
In (\ref{Hp}) $\mu$, $\omega_\rho$ and $\omega_z$  are, respectively, the reduced 
mass, oscillator frequencies in ($x^\prime$ and $y^\prime$) and $z^\prime$ directions respectively. $C$ and $D$ are 
strengths of the spin-orbit and $\vect{l\!\cdot\kern-1.5pt l}$ interaction.

The properly anti-symmetrical and normalized wavefunction of the Hamiltonian (\ref{H}) is given by
\begin{multline} 
\label{Psi}
|\,I,M\!> \,= \left [ 2I+1 \over 16 \pi^2 \right ]^{1/2}  
   \sum_{K=\Omega,\nu} C_{K,\nu} \left\{
   D_{MK}^I ( \theta ) \chir_{\Omega ,\nu}  \right.\\  
   + \left. (-1)^{I- 1/2} D_{M,-K}^I ( \theta )
   \chir_{-\Omega , \nu}  \right\}  \varphi_c  
\end{multline}
In (\ref{Psi}), $ D_{MK}^I$ is the eigenfunction of $\vect{I}^2$  with space and body fixed 
$z$-projections $M$ and $K$, respectively and $\theta$ is the three Euler angles.  
$\varphi_c$ is the normalized wavefunction of $H_c$.   $\chir_{\Omega ,\nu}$   
is the eigenfunction of the Nilsson Hamiltonian (\ref{Hp}) and is generated from the spherical shell model wavefunction $|j\,\Omega\!>$,  $\Omega$ being the 
$z$-projection of $j$ in the body-fixed system as follows:
\begin{equation}
   \chir_{\Omega , \nu} = \sum_j{\coefc_{j,\Omega,\nu} }\,
       |\, j \Omega \!> 
\end{equation}The index $\nu$  distinguishes among the Nilsson states of the 
same $\Omega$ originating from different orbitals at zero deformation.

The coefficients $\coefc_{j,\Omega,\nu}$  is obtained by diagonalizing 
(\ref{Hp}) and the band mixing coefficients $C_{K,\nu}$  are obtained by diagonalizing 
(\ref{H}) among all single particle and hole excited bands within a major shell. 
We have used three Cs for three distinctly different quantities: $C_{K,\nu}$s for
band mixing coefficients, $\coefc_{j,\Omega,\nu}$ for coefficients of the spherical shell
model wave functions to obtain Nilsson wavefunction, and $C$ for the strength of the spin-orbit 
coupling.
The coefficients $\coefc_{j,\Omega,\nu}$  are related to Nilsson's coefficients 
$\nilsona_{l,\Lambda}$ as 
follows (after suppressing the index $\nu$):
\begin{equation}
\label{cjomega}
\coefc_{j,\Omega} = \sum_{l,\Lambda} <\frac{1}{2} l \Lambda \Sigma | j \Omega> 
\nilsona_{l,\Lambda}
\end{equation}
The oscillator frequency at zero deformation, $\hbar\omega$, is set to be 
41 MeV/$A^{1/3}$, ``$A$'' being the mass number. The parameters of moment of 
inertia, $\mathscr A$ = $\hbar^2/2\mathscr I$, the deformation, $\beta$, and the strengths of the spin-orbit and orbit-orbit coupling along with 
the core-overlap or quenching factor $Q$ of the Coriolis coupling term 
($\vect{I\!\cdot\!j}$) in (\ref{H}) are   
 adjusted to reproduce the low lying observed level spectra. The band-head energy, E, in the model, is not a free parameter but obtained by summing appropriately over the Nilsson's eigenvalues, $E_{\Omega,\nu}$  and is given by
\begin{equation}
\label{E}
    E = \sum_\nu{1 \over 2} (1 + \mu / 2M)\, E_{\Omega,\nu}  -
        ( \mu / 4M)\,  <C \, \vect{l}\kern-3pt\cdot\kern-3pt\vect{s}
        + D \, \vect{l}\kern-2pt\cdot\kern-3pt\vect{l}>
\end{equation}
In (\ref{E}), $M$ is the mass of a nucleon.

\bigskip
\noindent{\it B. Super-allowed beta-decay rate}

\medskip
\noindent Under the assumption that the transition matrix $\vect{S}$ in beta-decay is 
energy independent, the comparative half-life, $ft$, is given by\cite{11}
\begin{equation}
\label{ft}
  ft =  {2 \pi^3 \hbar^7 c\, \ln 2  \over  (mc)^5 \, g^2 \, |S|^2 }
\end{equation}
In (\ref{ft}), $m$, $c$, and $g$ are, respectively, the rest mass of electron, 
the velocity of light in vacuum, and the coupling constant. And $f$ is given by 
the following expression:
\begin{equation}
\label{f}
f = \int_1^{W_0} F(Z,W) (W^2 - 1)^{1/2}\, (W_0 - W)^2 \, W\, dW
\end{equation}
In (\ref{f}), $W$ and $W_0$ are, respectively, the total and maximum energies of 
beta particle in the units of $mc^2$.  $F(Z,W)$ is the energy distribution 
function of beta-particle. Denoting the parities of the parent and the daughter 
states by $\Pi_P$ and $\Pi_D$, respectively and the magnitude of the orbital 
angular momentum of beta-particle by $l_\beta$, the parity conservation in the 
decay process is given by
\begin{displaymath}
\Pi_P = (-1)^{l_\beta} \Pi_D
\end{displaymath}
The matrix-element $|S|^2$ is usually expanded in terms of the ascending values 
of $l_\beta$. Thus,
\begin{displaymath}
   | S |^2  = | S(l_\beta = 0) |^2 + | S(l_\beta = 1) |^2
     + | S(l _\beta = 2) |^2   + \cdots
\end{displaymath}
The matrix elements associated with $l_\beta$ = 0, 1, 2\dots etc.\ are, 
respectively, termed as allowed, first forbidden, second forbidden, etc.

For allowed transition (\ref{ft}) can be rewritten as follows:
\begin{equation}
\label{ftallowed}
  ft = {F \over   < \vect{\bf 1}> ^2 + \:( g_A / g_V )^2
  <\!\boldsymbol\sigma\!\!>^2  }
\end{equation}

From the analysis of the decay of free neutron $( g_A / g_V )^2$,  the ratio of 
the coupling strength of the matrix elements of $<\!\vect{1}\!>$ and 
$<\!\boldsymbol\sigma\!\!>$, is determined to be (1.239 $\pm$ 0.011), and $F=$ 6150. This is the 
value used by Landford and Wildenthal\cite{5}.

A number of people, particularly Wilkinson\cite{12}, has pointed out that the 
ratio of the coupling constant and the value of $F$ in beta-decay from 
nuclei are likely to differ from the beta-decay rates of a nucleon in 
isolation. From the analysis of data, Wilkinson obtained   
and $( g_A / g_V )^2=$ 1.27 and 
$ F = (6162 \pm 14) (1-3.67 \times 10^{-4} Z + 1.30 \times 10^{-5} Z^2 )$, 
$Z$ being the atomic number of parent nucleus. In this paper, we shall present results using both sets of constants.   

The allowed transition between two iso-baric multiplets i.e., states having the 
same iso-spin $T$ but different $z$-component, $T_3$, are termed super-allowed 
transition. The matrix element of $<\!\vect{1}\!>$ between a parent state of 
total spin $I$, $z$-projection $M$, iso-spin $T$ and its $z$-component $T_3$, 
$|I(M)T(T_3)\!\!>$ to a daughter state of spin $I^\prime$, its $z$-projection $M^\prime$, 
iso-spin $T$ and its $z$-component $T^\prime_3$ is given by
\begin{eqnarray}
     < \vect{\bf 1}>  & \equiv &  <I^\prime ( M^\prime ) T^\prime
         ( T_3^\prime) | T_\pm | I (M) T ( T_3 )> \nonumber \\ 
     & = & \delta_{I I^\prime}  \:   \delta_{M M^\prime} \:
       \delta_{T_3^\prime , T_3 \pm 1}\, \sqrt{T(T+1)- T_3 T_3^\prime}  
\end{eqnarray}
Since the wavefunction (\ref{Psi}) is in body-fixed coordinate  
system ($r^\prime$), to determine the matrix-element of $<\!\sigma\!\!>$, 
$<\!\sigma\!\!>$ being an operator of rank one $O^1$, 
the latter is to be transformed from space-fixed ($r$) to 
body-fixed axis using the following transformation
\begin{equation}
\label{Sbs}
O^1_M (\vect{r} ) = \sum_{\mu}  D_{M \mu}^1 (\theta)\;
     O^1_{\mu } (\vect{r^\prime)}
\end{equation}
In (\ref{Sbs}), $M$ and $\mu$ are $z$-projections in space-fixed and body-fixed coordinates and ($\theta$) are three Euler angles.

Super-allowed transitions are characterized by the selection rule $\Delta I=0$ 
and $\Delta T=0$, and the spin and spatial part of the wavefunction of the parent 
and its daughter are the same. After the integration over the Euler angles, the matrix elements of Gamow-Teller transition becomes
\begin{eqnarray}
    <\!\boldsymbol\sigma\!\!> & \!\!\!\!= & \!\!\!\!{1 \over  \sqrt{I(I+1)}}  \sum_{K,\nu}{\!
     C_{K,\nu} \left[ K C_{K,\nu} \!
     <\!K,\nu | \sigma_0 | K,\nu\! > 
      + {1 \over 2} \sqrt{(I\!-\!K\!+1)(I\!+\!K)}\right. } \nonumber \\[5pt]
& &  \{ C_{K-1,\nu} + (-1)^{I-1/2}
      C_{1-K,\nu} \} <(K-1),\nu | \sigma_- | K,\nu> \nonumber \\[6pt]
& &  \left. + {1 \over 2} \sqrt{(I+K+1)(I-K)}
   C_{K+1,\nu} <(K+1),\nu | \sigma_+ | K,\nu > \right]
\end{eqnarray}
The prime over the summation indicates that the summation runs over all positive 
values of $K$ and $\nu$ i.e.\ $K=1/2$, 3/2, \dots  $I$.  The integration over the single particle matrix elements leads to the following expression for the Gamow-Teller moment for super-allowed transition:

\def\Ii{{I_1}}
\def\Mi{{M_1}}
\def\Ki{{K_1}}
\def\Iii{{I_2}}
\def\Mii{{M_2}}
\def\Kii{{K_2}}
\def\Iiii{{I_3}}
\def\Miii{{M_3}}
\def\Kiii{{K_3}}
\def\sumk{\sum_{k,\nu}}
\def\sumj{\sum_j} 

\begin{multline}
     \hskip-.4in<\!{\boldsymbol\sigma}\!\!>\;\; = 
      \;\; \sumk \left \{
     {K \over  \sqrt{I(I+1)}}   C_{K,\nu}^2
    \sumj \left ( {K \over j}  ( \coefc_{j,K,\nu}^+ )^2
   + \sqrt{1\!-\! {K^2 \over j^2} }\,
    \coefc_{j-1,K,\nu}^+   \coefc_{j,K,\nu}^+ \right.  \right. \\[5pt]
\left.  + \sqrt{1- {K^2 \over (j+1)^2}} \,
   \coefc_{j+1,K,\nu}^-  \,  \coefc_{j,K,\nu}^- 
   - {K \over j+1} ( \coefc_{j,K,\nu}^- )^2 \right ) \\[5pt]
+ \frac{1}{2} \sqrt{(I-K+1)(I+K) \over I(I+1)}\;  C_{K,\nu} 
  \left[ \sumj C_{K-1,\nu} \left(  {\coefc_{j,K,\nu}^+ \over j} \right. \right. \\[5pt]
\left[
  \sqrt{j(j+1)-K(K-1)}  \coefc_{j,K-1,\nu}^+
  - \sqrt{(j+K)(j+K-1)} \coefc_{j-1,K-1,\nu}^+  \right]  \\[5pt]
\hskip-.2in
  \left.  -  {\coefc_{j,K,\nu}^-  \over j+1}\!
  \left[ \sqrt{j(j\!+\!1)-K(K\!\!-\!1)}
  \coefc_{j,K-1,\nu}^- \!\!\! - \sqrt{(j\!-\!K\!\!+\!1)(j\!-\!K\!\!+\!2)}
  \coefc_{j+1,K-1, nu}^- \right] \right)      \\[5pt]
  + \;(-1)^{I+j-1} C_{1-K,\nu} \left(  {c^+_{j,K,\nu}  \over j} \right. 
   \left\{
      \sqrt{j(j+1)-K(K+1)} c^+_{j,K,\nu}  \right. \\[5pt] 
   \shoveright{ \left. 
   +\, \sqrt{(j+K)(j+K-1)} c^+_{j-1,K,\nu}  \right\} 
     -  {c^-_{j,K,\nu}  \over j+1} \quad}     \\[5pt]
  \left. \left. \left\{
  \sqrt{j(j+1)-K(K-1)} c^-_{j,K,\nu}
   + \sqrt{(j-K+1)(j-K+2)} c^-_{j+1,K,\nu} 
        \right\} \rule{0pt}{20pt}\right) \right]    \\[5pt]
  \shoveleft{}+ \frac{1}{2} \sqrt{(I+K+1)(I-K) \over I(I+1)} 
  C_{K+1,\nu} C_{K,\nu} \sumj \left(  {\coefc_{j,K,\nu}^+ \over j} \right. \\[5pt]
 \shoveright{}\left[  \sqrt{j(j+1)-K(K+1)} \coefc_{j,K+1,\nu}^+
  + \sqrt{(j-K)(j-K-1)} \coefc_{j-1,K+1,\nu}^+ \right]   
\nonumber
\end{multline}
\begin{multline}
    -  {\coefc_{j,K,\nu}^-  \over j+1} \left[  \sqrt{(j+K+1)(j+K+2)}
  \coefc_{j+1,K+1,\nu}^- \right. \\[5pt]
  \left.  \left.  \left.  
   + \sqrt{j(j+1)-K(K+1)} \coefc_{j,K+1,\nu}^- \right] \rule{0pt}{20pt} \right) \right\} 
\end{multline}
In the above, ${\large c}^+$ is the coefficient where $j=l+1$ and 
${\large c}^-$  when $j=l-1$. 	

The expression for beta decay in Nilsson model of the above Gamow-Teller moment 
can be obtained by setting $K=1$, and $C_{K,\nu}=1$ and, omitting the summation over $K$:
\begin{multline}
 \hskip-.35in <\!\boldsymbol\sigma\!\!> \;\; = \; \sqrt{I \over (I+1)}  \left[
                 \sum_j \left\{
                 {I \over j}\,  (c^+_{j,I})^2
               + \sqrt{1- {I^2 \over j^2}}\,  c^+_{j-1,I} c_{j,I} 
                       \right.  \right. \\[5pt] 
   \left.  + \sqrt{1- {I^2 \over (j+1)^2}}\,  c_{j+1,I}^-    c_{j,I}^-
   - {I \over j+1}\, ( c_{j,I}^- )^2 \right\} 
 + \, \delta_{I,1/2}\, (-1)^{j-1/2}  \\[5pt]
\left\{ \rule{0pt}{20pt}\right.
     {c^+_{j,I}\over j}  \left[  \sqrt{j(j+1)-I(I+1)} c^+_{j,I}
   + \sqrt{(j+I)(j+I-1)} c^+_{j-1,I}   \right]   \\
  - {c^-_{j,I}  \over j+1} \left.  \left[ \sqrt{j(j+1)-I(I-1)} c^-_{j,I}
   + \sqrt{(j-I+1)(j-I+2)} c^-_{j+1,I}
  \right] \rule{0pt}{20pt} \right\}
\end{multline}
The above expression can be rewritten in terms of Nilsson's\cite{7} coefficient 
$\displaystyle a_{l,\Lambda}$
 using (\ref{cjomega}). One, then, obtains the following expression for transitions between two Nilsson level
\begin{equation}
    \sigmagtm  = \sqrt{I \over I+1}  \sum_l 
                       ( a^2_{l, I-1/2} - a^2_{l,I+1/2} )   
                      \,\delta_{I,1/2} \,{2 \over \sqrt{3}} \,\Pi_\chi
                         \sum_l  a^2_{l,0}
\end{equation}
The above is identical to Kim's\cite{6} expression for allowed transition.  
$\Pi_\chi$ in the above equation is the parity of the wavefunction.

\section{Application and Discussion}
We present here calculations for the super allowed transition rates between the mirror pairs,  ($^{21}_{11}$Na,\, $^{21}_{10}$Ne),  
($^{23}_{12}$Mg,\, $^{23}_{11}$Na),  ($^{25}_{13}$Al,\, $^{25}_{12}$Mg), and  
($^{27}_{14}$Si,\, $^{27}_{13}$Al) and compare our results with the calculations of the Nilsson model i.e., without band mixing and the shell model.

The parameters of the model, $\mathscr A$ $=(\hbar / 2\mathscr I\:\,)$, $C$, $D$, $Q$, 
and $\beta$, have been determined from the fits to the observed low lying level 
spectra of these nuclei except for  $^{21}$Na and  $^{21}$Ne in \cite{4} and 
those for $^{21}$Na and  $^{21}$Ne in this paper. Calculated as well as observed 
level spectra up to approximately 3 MeV excitation energy for these nuclei are shown in Figs.~1 to 4 and the agreements are reasonable. As a further test to the model, the ground state magnetic and quadrupole moments of these nuclei have been calculated using the appropriate expressions for these quantities presented in 
\cite{9}. These calculations along with the observed moments are presented in 
Table~1 and the agreement is reasonable.

\begin{figure}
\noindent\begin{minipage}[t]{2.5in}
\includegraphics[bb=110 140 505 646,width=2.3in,height=2.95in,keepaspectratio]{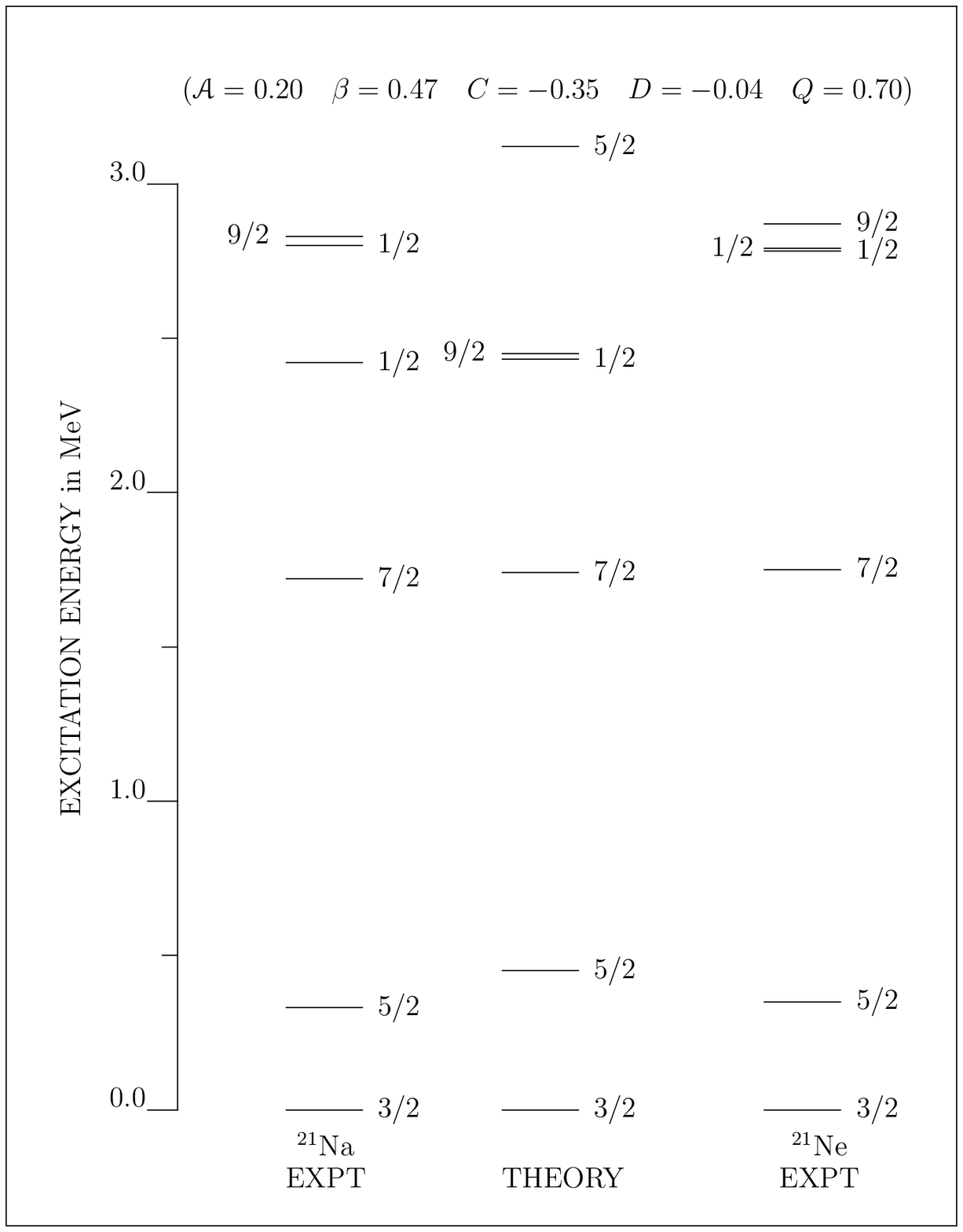}
\parbox{2.3in}{\caption{The Experimental level spectra of the mirror
nuclei $^{21}$Na  and  $^{21}$Ne  along
with the calculated spectrum shown in the middle.  The
parameters which produce the calculated spectrum
are shown on the top of the figure, $\mathscr A$ in MeV,
$C$ and $D$ in the unit of $\hbar \omega$.}\label{fig1}
}
\end{minipage}\begin{minipage}[t]{2.5in}
\includegraphics[bb=110 140 505 646,width=2.3in,height=2.95in,keepaspectratio]{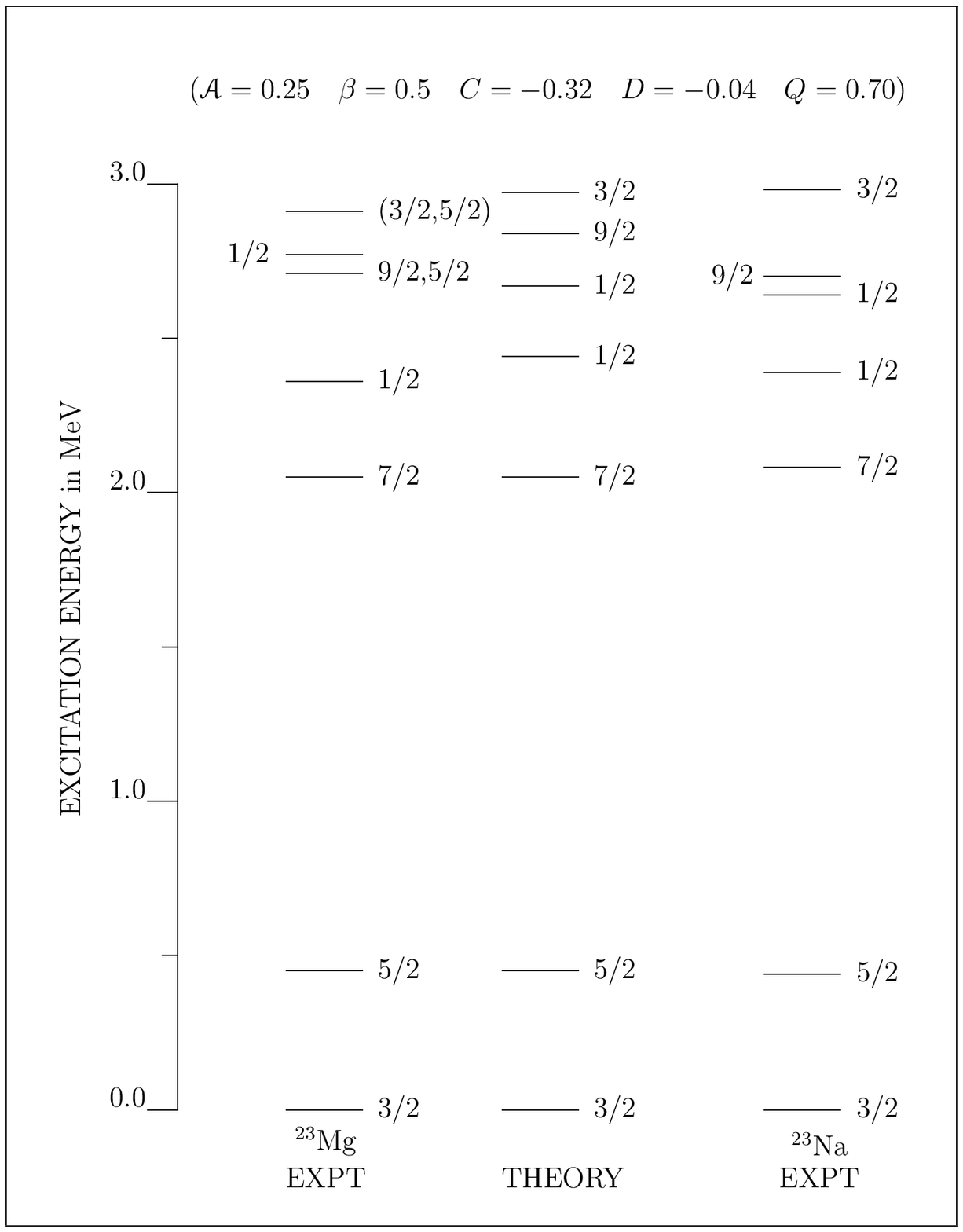}
\parbox{2.3in}{\caption{The Experimental level spectra of the mirror
pair $^{23}$Mg  and  $^{23}$Na.  The calculated spectrum is 
shown in the middle.  The parameters used for the calculated 
spectrum are shown at the top, in the same unit as in 
Figure~\ref{fig1}.}
}

\end{minipage}
\end{figure}

\begin{figure}
\begin{minipage}[t]{2.5in}

\includegraphics[bb=110 140 505 646,width=2.3in,height=2.95in,keepaspectratio]{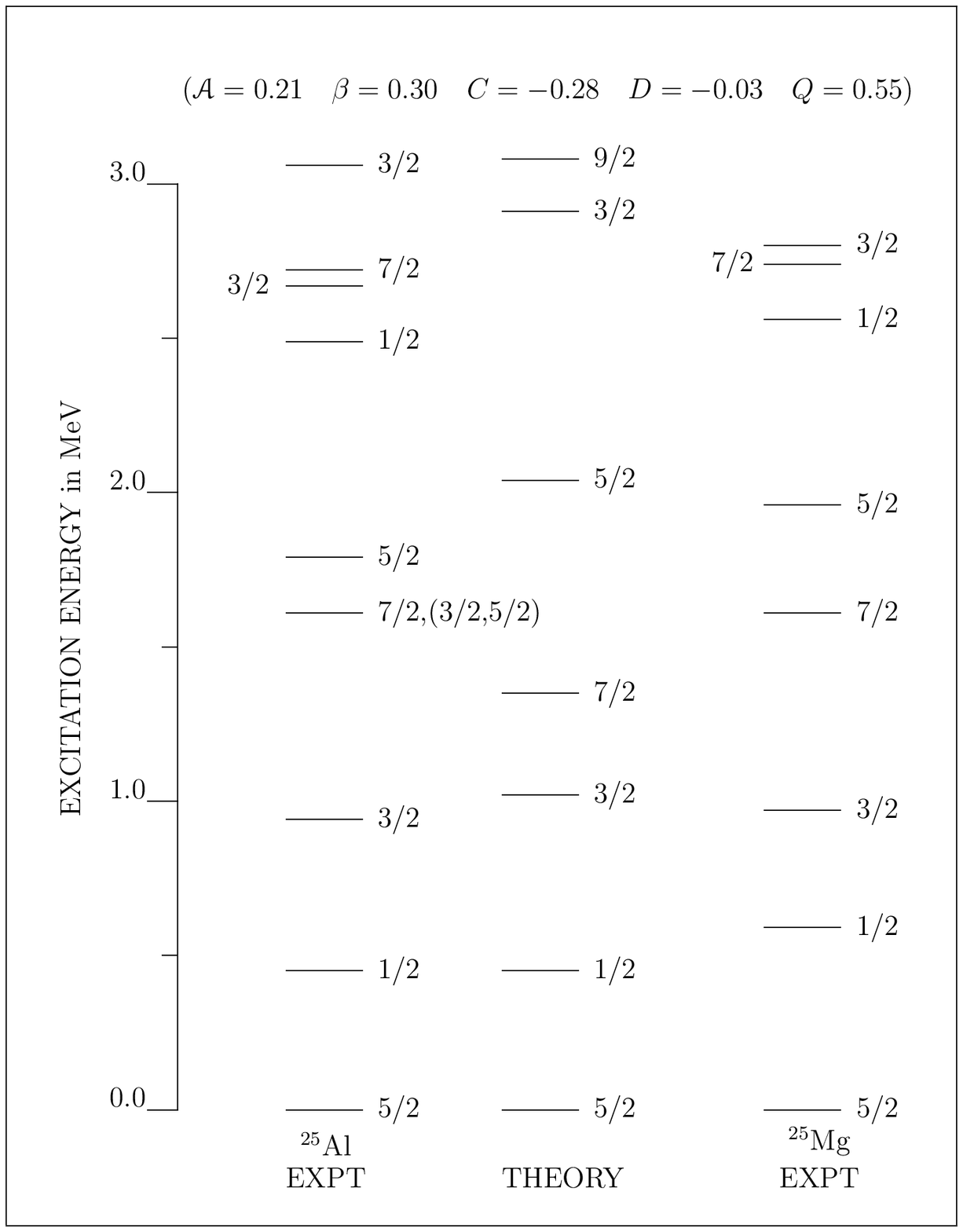}
\parbox{2.3in}{\caption{The Experimental level spectra of the mirror
pair $^{25}$Al  and  $^{25}$Mg.  The calculated spectrum is 
shown in the middle.  The parameters used for the calculated 
spectrum are shown at the top, in the same unit as in 
Figure~\ref{fig1}.}
}

\end{minipage}\begin{minipage}[t]{2.5in}

\includegraphics[bb=110 140 505 646,width=2.3in,height=2.95in,keepaspectratio]{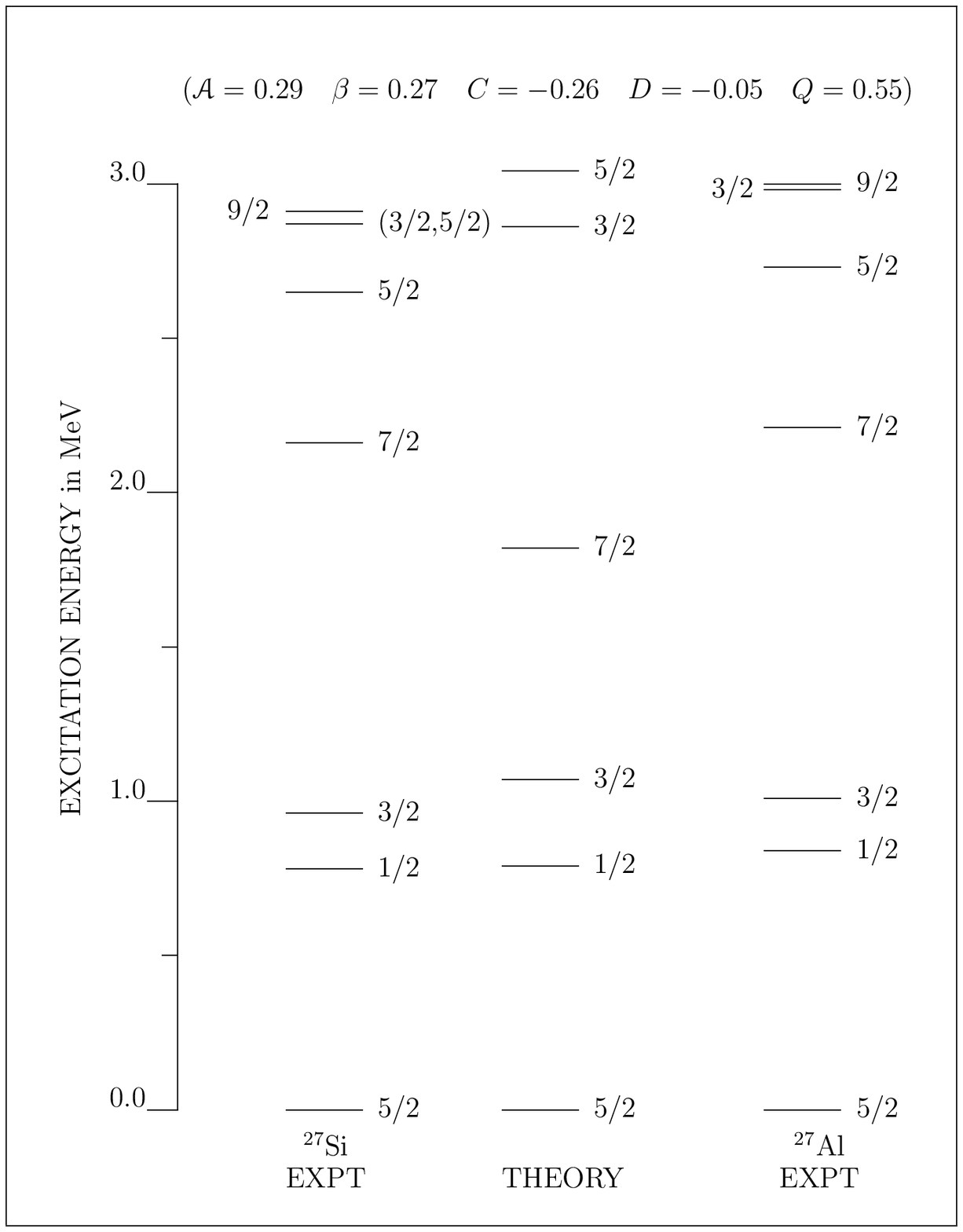}
\parbox{2.3in}{\caption{The Experimental level spectra of the mirror
pair $^{27}$Si  and  $^{27}$Al.  The calculated spectrum is 
shown in the middle.  The parameters used for the calculated 
spectrum are shown at the top, in the same unit as in 
Figure~\ref{fig1}.}
}
\end{minipage}
\end{figure}

\begin{table}
\caption{Calculated, noted as Cal, and experimental, noted as Exp, 
magnetic dipole, $\mu$, and electric quadrupole, $Q$, moments. 
The data are from \textit{National Nuclear Data Center, Database version of January 22, 2009}}

\vskip.25\baselineskip
\begin{tabular}{|c|r@{.}l|r@{.}l|c|r@{.}l|} \hline\hline 
\raisebox{-8pt}{NUCLEUS}  & \multicolumn{4}{c|}{$\mu$ in nm} & \multicolumn{3}{c|}{$Q$ in eb} \\ \cline{2-8} 
            & \multicolumn{2}{c|}{Cal} & \multicolumn{2}{c|}{Exp} & Cal
                                                       & \multicolumn{2}{c|}{Exp} \\ \hline & \multicolumn{2}{c|}{} & \multicolumn{2}{c|}{} & & \multicolumn{2}{c|}{} \\[-11pt]
$^{21}$Na  & 2&15     & 2&386                  & 0.075 & 0&05           \\[3pt]
$^{21}$Ne  & -0&564   & -0&662                 & 0.072 & 0&103           \\[3pt]
$^{23}$Mg  & -0&582   & 2&218                  & 0.100 & \multicolumn{2}{c|}{-}\\[3pt]
$^{23}$Na   & 2&172    & \multicolumn{2}{c|}{-}  & 0.084 & 0&101 \\[3pt]
$^{25}$Al   & 3&399    & 3&646                  & 0.101 & \multicolumn{2}{c|}{-}\\[3pt]
$^{25}$Mg   & -0&779   & -0&855                 & 0.112 & 0&20 \\[3pt]
$^{27}$Si   & -0&778   & \multicolumn{2}{c|}{-}  & 0.122 & \multicolumn{2}{c|}{-} \\[3pt]
$^{27}$Al   & 3&399    & 3&642                  & 0.092 & 0&140 \\ \hline\hline
\end{tabular}

\end{table}

In Table~2 we present calculated super-allowed ground state to ground state 
transition rates using both Wilkinson's and Landford and Wildenthal's\cite{5} parameters for $ft$ expression along with the measured one for these four pairs 
of mirror nuclei. The theoretical results are in good agreement with the observed 
one, particularly using Wilkinson's value for $ft$. The Fermi moment, 
$<\!\bf{1}\!>$, in these calculations has been set to 1.

\begin{table}
\caption{Calculated and observed super-allowed transitions 
between ground states 
of mirror nuclear pairs noted in the first column. Second to sixth columns 
represents the spin of initial and final states, calculated value of 
Gamow-Teller moment, $\log\!ft$ calculated with Wilkinson parameter, 
$\log\!ft$ calculated with Landford-Wildenthal's parameter and the 
observed $\log\!ft$  respectively. The data are from \textit{National 
Nuclear Data Center, Database version of January 22, 2009}.}

\vskip.25\baselineskip
\begin{tabular}{|r@{ $\rightarrow$ }l|r@{ $\rightarrow$ }l|r@{.}l|c|c|c|} \hline\hline
 \multicolumn{2}{|c|}{\raisebox{-8pt}{TRANSITION}} &  
\multicolumn{2}{c|}{\raisebox{-8pt}{$I_i \rightarrow I_f$}}  &  
\multicolumn{2}{c|}{\raisebox{-8pt}{$\sigmagtm$}}  & 
\multicolumn{3}{c|}{\raisebox{-2pt}{$\log ft$}}   \\ \cline{7-9}
\multicolumn{2}{|c|}{}&\multicolumn{2}{c|}{}&\multicolumn{2}{c|}{}& 
Cal(Wi) & Cal(LW) & Exp \\ \hline
$^{21}_{11}$Na&$^{21}_{10}$Ne \rule{0pt}{12pt}
             & $3/2^+$&$3/2^+$  & 0&553 & 3.65 & 3.63 & 3.61 \\[2pt] \hline 
$^{23}_{12}$Mg&$^{23}_{11}$Na \rule{0pt}{12pt}
             & $3/2^+$&$3/2^+$  & 0&443 & 3.65 & 3.63 & 3.67 \\[2pt] \hline
$^{25}_{13}$Al&$^{25}_{12}$Mg  \rule{0pt}{12pt}
             & $5/2^+$&$5/2^+$  & 0&731 & 3.56 & 3.54 & 3.57 \\[2pt] \hline
$^{27}_{14}$Si&$^{27}_{13}$Al \rule{0pt}{12pt}
             & $5/2^+$&$5/2^+$  & 0&692 &  3.58 & 3.51 & 3.62 \\[2pt] \hline 
\end{tabular}

\end{table}

Kim's calculations using the Nilsson's model, i.e., without taking into consideration the band mixing caused by the Coriolis coupling, yields an $ft$ 
value of 3.49 for  $^{25}_{13}$Al$ \rightarrow ^{25}_{12}$Mg and  
$^{27}_{14}$Si$ \rightarrow ^{27}_{13}$Al transitions, which deviates 
significantly from the observed value of 3.6 in both cases. Thus, the band 
mixing affects significantly the calculations of super-allowed transition.

As is well known, the elementary single particle shell model can not predict the 
correct ground state spins of  $^{21}$Na,  $^{21}$Ne,  $^{23}$Mg, and  $^{23}$Na 
and hence, can not account for the observed \,$\log\!ft$\, values. The 
consideration of residual interaction in the shell model calculations 
rectifies the situation. Landford and Wildenthal have calculated the level 
spectra of many of these nuclei using  $^{16}$O as a core and configuration 
mixture among nucleons outside the core. Their calculated \,$\log\!ft$\, values 
for super allowed transitions  
$^{21}_{11}$Na$\rightarrow ^{21}_{10}$Ne,\,  
$^{23}_{12}$Mg$\rightarrow ^{23}_{11}$Na,\,  
  and\,  
$^{27}_{14}$Si$\rightarrow ^{27}_{13}$Al are, respectively, 3.56, 3.68, and 3.51 
(they did not report  $^{25}$Al$\rightarrow ^{25}$Mg rate), which are close to 
our and observed values.

\section{Conclusion}

\noindent We have derived here the expression for super-allowed transition for the 
Coriolis coupling model of Malik and Scholz, which are valid for transition between 
any pair of mirror nuclei. The application to four pairs of $1d_{1/2}$-nuclei 
indicate that the band mixing affects the rate. Whereas calculations based on the 
Nilsson model are at variance with the data, calculations presented here and based on the shell model with configuration mixing presented in \cite{5} can account for 
the observed rates.

\end{document}